# Enhancing User Engagement in Large-Scale Social Annotation Platforms: Community-Based Design Interventions and Implications for Large Language Models (LLMs)

JUMANA ALMAHMOUD, Computer Science and Artificial Intelligence Laboratory (CSAIL), Massachusetts Institute of Technology, USA
MARC FACCIOTTI, Biomedical Engineering, University of California, Davis, USA
MICHELE IGO, Biomedical Engineering, University of California, Davis, USA
KAMALI SRIPATHI, Genome Center, University of California, Davis, USA
DAVID KARGER, Computer Science and Artificial Intelligence Laboratory (CSAIL), Massachusetts Institute of Technology, USA

Social annotation platforms enable student engagement by integrating discussions directly into course materials. However, in large online courses, the sheer volume of comments can overwhelm students and impede learning. This paper investigates community-based design interventions on a social annotation platform (NB) to address this challenge and foster more meaningful online educational discussions. By examining student preferences and reactions to different curation strategies, this research aims to optimize the utility of social annotations in educational contexts. A key emphasis is placed on how the visibility of comments shapes group interactions, guides conversational flows, and enriches learning experiences.
The study combined iterative design and development with two large-scale experiments to create and refine comment curation strategies, involving thousands of students. The study introduced specific features of the platform, such as targeted comment visibility controls, which demonstrably improved peer interactions and reduced discussion overload. These findings inform the design of next-generation social annotation systems and highlight opportunities to integrate Large Language Models (LLMs) for key activities like summarizing annotations, improving clarity in student writing, and assisting instructors with efficient comment curation.

## 1 INTRODUCTION

Social annotation (SA) tools, which situate discussions in the margins of online learning resources (Figure 1), can significantly enhance interactive and collaborative learning experiences [6, 9, 26]. SA tools enable students to discuss and analyze course materials, fostering a richer engagement with educational content. However, in large classes, the sheer volume of annotations can overwhelm students, deterring engagement and obscuring valuable insights [9, 20, 23].

Traditionally, educators have managed scale by segmenting classes into smaller *sections*. Yet, the inherent value of scale in online discussions must be considered, as it offers diverse perspectives and rapid peer feedback [5]. Our study proposes a nuanced approach that leverages comment curation to maintain the benefits of scale while mitigating its challenges. By implementing various curation strategies, we explore how to optimize social annotations for educational engagement,

Authors' addresses:, jumanam@mit.edu; Jumana Almahmoud, Computer Science and Artificial Intelligence Laboratory (CSAIL), Massachusetts Institute of Technology, Cambridge, MA, USA, 02139; Marc Facciotti, Biomedical Engineering, University of California, Davis, Davis, USA, mtfacciotti@ucdavis.edu; Michele Igo, Biomedical Engineering, University of California, Davis, Davis, USA, mmigo@ucdavis.edu; Kamali Sripathi, Genome Center, University of California, Davis, Davis, USA, ksripathi@ucdavis.edu; David Karger, Computer Science and Artificial Intelligence Laboratory (CSAIL), Massachusetts Institute of Technology, Cambridge, MA, USA, karger@mit.edu.

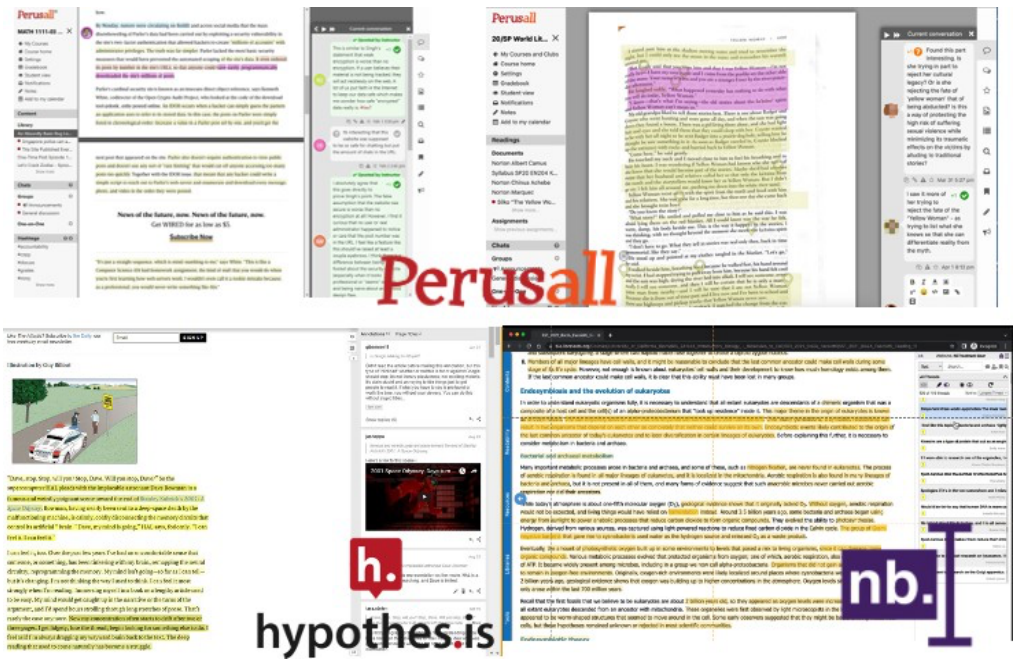

Fig. 1. Example of SA tools. Discussions usually happen on the right margins, and the document is filled with highlights associated with that discussion. In highly active classes this tend to be overwhelming.

focusing on comment visibility and the presentation of curated content on student interaction. We investigate how highlighting specific comments, the balance between focusing on the content and the social interaction, and making comments more visible versus when instructors endorse them can affect student engagement.

To manage and curate comments effectively in SA platforms, our research explores different levels of visibility that leverage the unique affordances of social annotations. By utilizing the margins, the document itself, and the annotations, we aim to engage students more effectively with the reading material and with each other. A key aspect of this investigation is structured around central questions concerning the methods of curation, effective mechanisms for managing annotations, presentation of curated content, and the impact of comment salience on student engagement.

This study unfolds within the framework of an introductory biology course conducted across the Winter and Fall terms of 2023 at a large public university. Through experiments with 1,323 students on a modified version of the open source SA platform Nota bene (NB) [1], we assess the impact of various curation methods on students' engagement preferences, behaviors, and overall learning experience, aiming to enhance the educational value of social annotation in large-scale learning environments.

Our synthesis reveals that strategies such as increasing the visbility of instructor-endorsed comments or enabling students to control the visibility of their contributions elevate students' engagement. These strategies highlight the importance of perceived authority and relevance in driving interactions and reflect a preference among students for a learning environment that values and seeks their input. Moreover, the introduction of response-inviting prompts and marginal cues enhances students' engagement with the material and each other.

---

[1]https://github.com/haystack/nb

Our research concludes that curated engagement strategies, particularly those that manipulate visibility levels through social annotations, can significantly enhance the educational utility of SA tools in large-class settings. By aligning these strategies with students' engagement preferences and incorporating thoughtful design elements, we can overcome the challenges posed by scale and create a more effective learning environment.

Fig. 2. Enhanced interface of NB for Study 1 with curated spotlit annotations (1) showing instructor endorsements and author follows, sidebar (2) displaying filtered comments with on-demand expansion.

## 2 RELATED WORK

### 2.1 Student Engagement in Social Annotation in Large-Scale Online Classrooms

Social Annotation (SA) tools transform the reading experience from passive consumption to an engaging and interactive process. By enabling collaborative commenting on digital content, SA tools empower learners to actively participate in the educational discourse, thereby contributing to a deeper understanding and more meaningful learning experiences [6, 9, 26]. Benefits of SA include improved reading comprehension, enhanced critical thinking, and increased student confidence. Exposure to diverse perspectives aids in identifying knowledge gaps, while real-time peer feedback supports a dynamic learning process [7].

However, implementing SA tools in large-scale online classrooms introduces significant challenges. In high-enrollment classes, the sheer volume of annotations can lead to information overload, potentially diminishing the effectiveness of collaborative learning and peer review [5, 13]. In addition, the abundance of discussion threads can cause valuable contributions to become lost, thereby reducing the overall quality of the learning experience [21]. Developing effective curation strategies that guide students toward high-value discussions is crucial to addressing these issues, thus enhancing the overall educational experience [20, 23].

### 2.2 Visibility of Comments in Online Discussion Platforms

Online discussion platforms face the challenge of balancing the visibility and relevance of comments, which significantly influences user behavior and engagement. Backstrom et al.'s research on Facebook's News Feed algorithm sheds light on how visibility due to algorithmic sorting impacts user interaction [2]. Though their study centers on social media posts, the underlying principles about the relationship between content visibility and user engagement are universally relevant, including in the context of educational discussion forums. Bernstein et al.'s exploration of the 'invisible audience' in social networks contributes to understanding how both visible and invisible audiences affect online behavior [4]. These studies collectively offer important perspectives on the dynamics of visibility in online communities, though they do not address the specific aspect of elevating certain comments.

The concept of enhancing the quality of online discussions through the strategic visibility of comments is further examined in Ziegele et al.'s research in online news forums [25]. Their findings suggest that platforms that spotlight high-quality comments can create an environment that encourages users to produce thoughtful and insightful contributions. This approach not only rewards quality but also sets a standard for discourse, fostering a self-regulating community. Additionally, Gillespie's discussion on algorithmic curation practices highlights the significant role these algorithms play in determining content visibility on various platforms, including discussion forums [11]. The way comments are algorithmically ranked can profoundly influence their visibility and, consequently, the direction and quality of the conversations that ensue. However, algorithms used to curate online discussions often introduce biases and marginalization and create 'filter bubbles' potentially silencing diverse voices [3].

### 2.3 Comment Curation in Online Discussion Forums

The practice of comment curation in online discussion forums is pivotal in shaping the nature and quality of discourse. Curation involves prioritizing comments that exhibit valuable qualities such as relevance, thoughtfulness, or diverse perspectives. This process is instrumental in structuring conversations that are not only informative but also conducive to a productive learning environment. Cliff Lampe and Paul Resnick explore distributed moderation in large online conversation spaces like Slashdot, revealing how community-based curation can influence the dynamics of discussion [16]. The study offers a perspective on how distributed curation can be employed in educational settings, potentially enhancing student engagement and the depth of discourse. Sukumaran et al. highlight the importance of normative influences on online participation. Their research points out that thoughtful participation in online forums is often guided by the perceived norms and expectations set by the community [22]. This insight is crucial for understanding how comment curation can shape student behavior and engagement in online learning platforms.

Further, Lampe et al.'s work on crowdsourcing civility in online forums underscores the potential of community-driven curation in fostering constructive and respectful discussions [17]. This approach to moderation can be particularly effective in educational contexts, where maintaining a respectful and inclusive environment is essential. However, previous work indicates that peer moderation can lead to underprovision, a situation where too many people rely on others to contribute without contributing themselves [10]. This issue can significantly reduce the effectiveness of collaborative discussions.

*2.3.1 Comment Curation in Education.* Scaling online educational discussions to accommodate a large number of learners presents several challenges, including group creation, activity monitoring, collaborative coordination, and attention management [5]. Various strategies have been implemented to address these. For instance, FutureLearn, a MOOC platform, simplifies discussion threads by dividing students into adaptive smaller groups "buses", thus ensuring learners don't get overwhelmed and facilitating focused group discussions [8]. Another approach is using scripts in MOOCs to direct learners into expertise-based sections [12]. These initiatives represent the blend of technology and pedagogy aiming to handle large-scale issues, but they all minimize the scale of discussion into smaller groups which leads to delayed feedback and limited peer support [15]. We know from the literature that an immense value comes with scale [5]. The vastness of online discussion forums in education, such as the ones on MOOCs, enabled prompt peer feedback [15] and reduced wait times for peer reviews [8]. Emerging research has begun to illuminate the role of comment curation in fostering self-directed learning. This aspect of curation is particularly relevant in online education, where it has the potential to transform the learning experience, making it more engaging and effective [14].

Instructors play a pivotal role in curating conversations to facilitate engagement. By highlighting instructionally valuable threads, they can guide students towards more productive discussions, fostering a more collaborative learning experience [19]. This selective elevation of comments and discussions serves not just to direct attention but also to shape the discourse in a manner conducive to learning [1].

## 3 BACKGROUND AND NEED-FINDING STUDY

Our work was motivated by feedback from instructor collaborators who identified a significant challenge in using SA tools in their classroom, such as NB: the sheer volume of student-generated comments. While these comments could deepen student engagement and understanding, they often led to information overload, diluting the quality of discourse and posing a management challenge for students. Furthermore, it was observed that a portion of the commentary was driven by the incentive of grades rather than genuine engagement, adding to the 'noise' and obscuring valuable discussions. To address the challenges associated with using social annotation platforms in large online courses, we conducted a need-finding study with 30 students who have used the platform NB. These semi-structured interviews, each lasting approximately 30 minutes, were conducted online to accommodate the students' diverse schedules and geographic locations. Our participant group consisted of undergraduate students studying a biology class. The study revealed several key issues that students face when interacting with the platform. Firstly, a number of students (n=14) reported difficulties in finding relevant discussion areas to comment on, particularly if they participated later in the course timeline. This timing issue impacts student engagement, as latecomers often find themselves either echoing previous comments or struggling to contribute meaningfully. Students expressed a need for mechanisms to guide their attention to areas where they can actively participate and contribute, enhancing their engagement with each other and the reading. Secondly, more than half of the participants (n=18) confessed that time pressures and difficulties understanding the course material frequently led to rushed or low-quality comments. A method to distinguish these hurried contributions from more thoughtful ones, aiming to foster more productive and less overwhelming discussions, is clearly needed. Furthermore, while some students (n=9) resorted to seeking clarification from friends outside the platform due to a lack of timely responses within NB, a substantial majority (n=25) greatly valued the direct involvement of instructors in the discussions. This preference underscores the importance of striking a balance between instructor-led interventions and student-initiated discussions, which could potentially enhance the dynamics of the platform and improve student experiences. Lastly, the observation that many students (n=13)

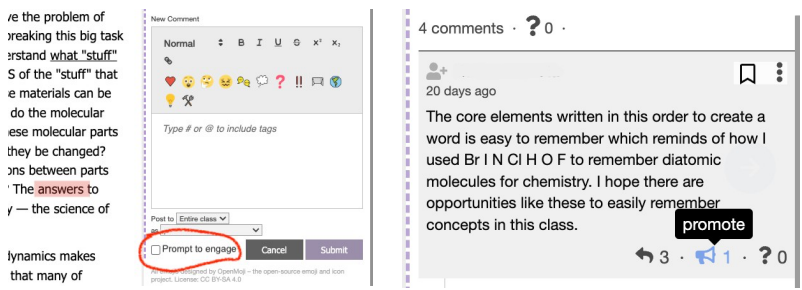

(a) Prompt when composing a comment.

(b) Prompt for published comments (Study 2)

Fig. 3. Study 2 used "Promote" as a prompt. Study 3 used "Response Wanted."

find it easier to respond to existing comments than initiate new ones underscores the inherent challenge in fostering proactive engagement. This suggests a need for strategies that encourage ongoing participation. These findings lay the groundwork for our subsequent research, effectively framing our questions and studies to directly address students' specific needs and behaviors by modifying the open-source SA tool NB.

## 4 DESIGN

Our need-finding study revealed several key challenges faced by students using social annotation tools in large courses. Informed by these findings, we designed two user studies (Study 1 and Study 2) to explore the effectiveness of comment curation strategies in addressing student needs and fostering meaningful engagement with their peers and learning material. Both studies employed a mixed-methods approach within the same introductory biology course offered at a large public university. The student population consisted of undergraduates with similar demographics across both studies, and the same instructors facilitated the course and participated in the curation tasks. Additionally, both studies utilized the same set of 26 course readings with consistent requirements for student annotation (3 mandatory comments per reading). However, each study explored distinct curation strategies to investigate their impact on student engagement and learning outcomes. We will delve into the specific designs, methodologies, and research questions of Study 1 and Study 2 in the following sections.

### 4.1 Strategy 1: Content-based Curation

Our first user study investigated the effectiveness of "content-based curation" in addressing student needs and promoting meaningful engagement. This strategy aimed to guide students towards valuable contributions by highlighting specific comments within the annotation landscape. In the context of a large class brimming with comments, our primary design objective was to create differentiated levels of visibility. This was achieved by implementing *spotlights*, a design element that amplifies the visibility of comments by moving them from the sidebar into the content area adjacent to the annotations [1]. We refined the concept of spotlights by labeling them based on the type of comment—whether instructor-endorsed, questions, or comments from a user's following list (Figure 2). This strategic categorization aimed to highlight content from knowledgeable figures (instructors), foster peer assistance by showcasing questions, and encourage social interactions by emphasizing content from followed users. Additionally, users were given the control to manage their engagement by following or unfollowing others. The sidebar had a filter to display the most recent comments alongside the curated content, ensuring broader visibility beyond the prioritized list.

*4.1.1 Research Questions.* This study was guided by several key research questions:

(1) Does the Type of Spotlighted Comment Affect Student Engagement? We aimed to understand if highlighting different types of comments (instructor-endorsed, questions, peer comments) influenced student interaction with the annotations.
(2) Are Students More Likely to Initiate Replies to Certain Types of Comments? We investigated whether specific spotlighted comments (e.g., questions) encouraged students to initiate new discussion threads.
(3) How Does the Preference for Content-Driven Engagement Over Personal Recognition Influence the Social Dynamics in Large Online Classes? We explored how highlighting content-based interactions (such as questions) compared to social recognition (followed users) might impact student behavior and class discussions.

*4.1.2 Methods.* The study was conducted during the Winter 2023 semester within an introductory biology course at a large public university. We employed a within-subjects design with a participant pool of 1,135 undergraduate students. Each reading passage featured a maximum of 12 spotlighted comments, divided equally among the three categories (instructor-endorsed, questions, or comments from a user's following list). The comments within each category were randomized to avoid any bias.

To measure student engagement, we analyzed usage logs, focusing on clicks and replies made on spotlighted comments. Additionally, we conducted semi-structured interviews with a subset of 25 students. Thematic analysis of these interviews provided insights into student motivations and experiences with the content-based curation strategy. Finally, we utilized Repeated Measures ANOVA to statistically analyze engagement data based on clicks and replies across the different highlighted comment categories.

## 4.2 Strategy 2: Curation Prompts

Building upon the findings of Study 1, our second user study explored the concept of "curation prompts" aimed at fostering genuine engagement and participation beyond mandatory contributions, which often lead to information overload. This approach focused on empowering students by providing them with agency over their participation in social annotations.

We introduced new features to encourage students to share their thoughts with intentionality. The first prompt introduced was "Promote," which allowed students to highlight their comments for increased visibility (Figure 3). Drawing on insights from the initial study, we recognized the importance of maintaining instructor-endorsed comments in the curated selection. However, we also aimed to address the diverse motivations behind student engagement. A dual-level visibility mechanism was introduced to cater to students seeking a less cluttered experience and those desiring wider exposure for their contributions.

By default, annotations were hidden behind filters that could be cleared easily by students, with the highest level of visibility reserved for instructor-endorsed comments through page spotlights. Other comments could gain visibility in the sidebar if students opted to "promote" them. This term was chosen to signify the elevated visibility and broader audience reach these annotations would achieve. This design empowered students to decide the prominence of their comments, fostering a tailored and interactive learning environment that catered to individual preferences.

Recognizing the potential shift in intentionality with the "Promote" option, we further refined the prompt to "Response Wanted" in a subsequent iteration (Figure 3). This subtle change aimed to encourage interaction without overtly emphasizing visibility, instead encouraging students to signal their openness to further engagement on their comments.

*4.2.1 Research Questions.* Our research questions in this study focused on the impact of these prompts:

(1) How does the change from visibility-focused to response-inviting prompts impact user engagement in online annotations? We aimed to understand how the shift in terminology from "Promote" to "Response Wanted" influenced student interaction with the annotation features.
(2) How does the instructor's assessment of student annotations compare to the students' own assessment of their comments? We investigated any discrepancies between instructor-endorsed comments and student-perceived value of their own contributions.

*4.2.2 Methods.* The study was conducted during the Fall 2023 semester with a participation pool of 188 students in an introductory biology course. We employed a within-subjects design, allowing students to experiment with both promoting and not promoting their comments for visibility. T-tests were used to compare the average number of clicks and replies received on comments that students chose to promote in the default view.

To gain deeper insights into student behavior and experiences with the curation prompts, we utilized a combination of qualitative and quantitative methods. Semi-structured interviews were conducted with a subset of 20 students, and a survey achieved a 75% response rate, gathering data on student promotion habits and experiences with the curation system. Additionally, to study the quality of potential promoted comments, an instructor-led evaluation process was implemented. A random sample of student annotations was assessed based on criteria developed by the instructors: successful resolution of confusion, contribution to building the discussion, insightfulness, contextual relevance, and potential helpfulness to other students.

## 5 RESULTS
## 5.1 Study 1 Results

Our study, involving 1135 undergraduate students in an introductory biology class, focused on analyzing engagement patterns with spotlighted comments in NB. We examined metrics, including clicks, replies, likes, and follow/unfollow actions, to gain insights into initial student interactions. Overall, students appreciated the idea of labeled spotlights and how accessible they are, as expressed by this student: "I think that's cool. And I think that having it in the relevant place is really great. I love that. It's labeled clearly of like, why, it's important and why it's there."

In total there were 12,076 clicks, with an average of 22 comments made per student. An in-depth look at the 4,161 clicks initiated on the spotlights on the document revealed FOLLOW comments as the most clicked category, garnering 1,305 clicks (31.36%), followed by ENDORSED comments (1,090 clicks; 26.20%), QUESTION comments (1,067 clicks; 25.64%), and GENERAL comments (699 clicks; 16.80%). Replies initiated on spotlights totaled 486, with QUESTION comments leading significantly at 198 replies (40.74%), indicating heightened engagement for clarification or discussion. GENERAL, FOLLOW, and ENDORSED comments followed with 101 (20.78%), 96 (19.75%), and 91 (18.72%) replies, respectively.

*5.1.1 Does the Type of Spotlighted Comment Affect Student Engagement?* A Repeated Measures ANOVA was conducted to examine the effect of the type of spotlighted comment (ENDORSED, FOLLOW, GENERAL, QUESTION) on student engagement, as measured by clicks and replies. The analysis revealed a significant effect of comment type on clicks, $F(3, 2607) = 22.14$, $p < .001$, and on replies, $F(3, 2607) = 18.48$, $p < .001$.

Post hoc comparisons using the Tukey HSD test indicated significant differences in student engagement across comment types. The results for clicks revealed that QUESTION comments received significantly fewer clicks than FOLLOW comments (meandiff = -0.6816, $p$ = 0.0097), but the difference between QUESTION and ENDORSED was not significant. Our interviews uncovered that FOLLOW received the highest clicks since students clicked on these spotlights to access the comments to either follow or unfollow (as opposed to them attracting genuine engagement), as we will elaborate shortly.

Students found instructor-endorsed comments to help direct their reading and understanding of what was important or expected in the discussion. The clear labeling and relevance of these comments were appreciated, as they guided without limiting access to other contributions. Students also indicated that these instructor-endorsed comments set a model of the instructor's expectations regarding comments. Despite their popularity, they did not necessarily translate into many replies, as we will see in the following section. Students view endorsed comments as definitive or clarifying, leaving little room for further inquiry or discussion, leading them to feel that additional replies were unnecessary.

*5.1.2 Are Students More Likely to Initiate Replies to Certain Types of Comments?* The analysis for replies indicated that QUESTION comments received significantly more replies than ENDORSED (meandiff = 0.123, $p$ < 0.001), and FOLLOW (meandiff = 0.1172, $p$ < 0.001) comments. No significant difference in replies was observed between ENDORSED and FOLLOW comments.

Questions posed by peers were a significant point of engagement for many students. They preferred engaging with questions that either aligned with their own confusion or offered a deeper insight into the subject matter. As one student expressed: "Most of the ones [spotlights] I've clicked on were questions, and that's been helpful for me because usually, I reply to those questions. " However, many students (17) indicated that some of these questions are so simple and are there to fulfill the requirements. While they are mostly helpful, some are just noise.

*5.1.3 How does the preference for content-driven engagement over personal recognition influence the social dynamics in large online classes?* Regarding social dynamics on the platform, there were 190 instances where users chose to follow each other, contrasted with a notably higher 1,410 instances of unfollowing. A few students (4 out of 25) appreciated the social aspect of the following feature and tended to follow peers based on personal recognition from the classroom or discussion groups. One student was interested in following individuals who consistently provided valuable input during class discussions. This was further supported by another student's comment on being more excited to engage with students that she recognized who "put more effort into their comments" on NB.

However, the more common theme was the focus on the content of comments over the commenter's identity. One student noted, "I look at just the content and not really worry about who's posting what... interacting with the information and not the person behind it." This person pointed out that they would unfollow authors if they see comments that do not add value to them by clicking on the spotlight and unfollowing the author, regardless of the author's identity. This indicates a preference for content-driven engagement, particularly in large classes with limited personal recognition, as another student expressed when we asked him about recognizing and following other students, "It is a really big class. So not really, you know." Another student said, "I don't know how reliable that one person is. So I kind of like to look at variety of people's comments. If I can find a variety of consensus." She explained that she would unfollow as she would rather get spotlights based on content than a social following list.

## 5.2 Study 2 Results

After analyzing the dataset, we found that 6,661 annotations were made by users of which are 4,470 replies to various comments. Furthermore, 1105 comments were flagged to invite responses. The dataset also recorded 64,923 clicks.

*5.2.1 How does the change from visibility-focused to response-inviting prompts impact user engagement in online annotations?* Changing the wording from "Promote" to "Response wanted," increased the percentage of shared annotations on the default view from 9% to 16.59% out of 6,661 unique annotations. Statistical analysis showed a significant increase in clicks ($t = 26.438$, $p \approx 3.49 \times 10^{-68}$) and replies ($t = 15.872$, $p \approx 2.87 \times 10^{-37}$).

Based on discussions with students and our analysis of qualitative data, about 73% of students expressed reluctance to promote comments, primarily due to a lack of expertise or confidence in their judgment. They voiced concerns about overloading their peers' views and imposing their choices on others. In response to these issues, students suggested that the platform's prompts should be modified. Rather than focusing on the promotion or visibility of comments, they recommended using prompts that encourage specific actions such as 'find answers' or 'discuss.' They believed that this approach would facilitate more meaningful interaction by reducing the pressure of influencing content visibility.

*5.2.2 How does the instructor's assessment of student annotations compare to the students' own assessment of their comments?* According to the students we interviewed, they feel it is tough to judge their own comments and sometimes feel their comments are not worth promoting. In the results of a qualitative assessment of 53 randomly selected student comments performed by the instructor, we found that 55% (29 out of 53) of the annotations were deemed worth promoting. Of these, 57% (23 out of 40) were annotations students didn't promote, whereas 46% (6 out of 13) were annotations promoted by students. Within the subset of student-promoted annotations, 86% (6 out of 7) were considered worth promoting by the instructor. Furthermore, from the annotations not promoted by students, 58% (23 out of 40) were evaluated as worth promoting by the instructor.

## 6 DISCUSSION

In this section, we discuss the implications of our study findings on student engagement patterns with spotlighted comments in online discussion platforms. We explore the influence of different types of comments, the visibility of confusion and simplicity, and the taxonomy for curating comments based on instructor and student perspectives.

### 6.1 Affordances on SA Tools to Manage Scale and Improve Engagement

To effectively manage the scale of interaction without overwhelming participants, our designs use a strategy of multilayered visibility. This approach allows for annotations to be displayed with varied salience, making critical insights more accessible while keeping the discussion landscape organized and navigable. In our designs, annotations that instructors endorse are spotlighted directly in the text. This method guides the learning focus and anchors discussions around key concepts and ideas considered crucial by educators. Spotlighting acts as a navigational aid within large volumes of text, directing students' attention to areas with significant educational importance.

Furthermore, we empower students by enabling them to elevate certain annotations on the sidebar. This feature is designed to stimulate peer discussion and promote critical thinking by highlighting student-selected comments that offer new insights or present challenging viewpoints. Elevating these annotations ensures that significant student contributions are visible, enhancing peer-to-peer learning and engagement. These features of our SA tools are specifically designed to manage the vast number of interactions typical in large classes and to enhance the quality of those interactions. By providing both students and instructors with the ability to control the visibility and salience of annotations, our tools create an environment where meaningful academic discourse can thrive. The balanced dynamic of instructor-endorsed and student-curated annotations supports a learning process where authoritative guidance and independent critical thinking coexist, enhancing the overall educational experience in large-scale settings.

## 6.2 Enhancing Engagement Through Curation in Online Discussions: Strategies and Considerations

The challenge of fostering meaningful engagement in high-enrollment classes, where the volume of participation often overshadows the quality of discourse, is central to our investigation. This dilemma is exacerbated by the requirement for students to post comments for grading purposes, leading to a proliferation of low-quality contributions. Echoing Wise et al.[24], we recognize that authentic participation in online discussions extends beyond mere posting. Our endeavor seeks to elevate the caliber of interactions within these forums. We have adopted various strategies to enhance engagement and curate comments in large classes, such as spotlighting instructor endorsements, which assign high visibility to the endorsed comments. We also implemented a system that allows students to increase the visibility of their thoughtful contributions. Comments expressing confusion or seeking clarification were valued for their relatability, fostering a supportive environment where students felt comfortable sharing and addressing misunderstandings collectively.

A more in-depth exploration of these mechanisms' impact and endorsement strategies in various learning environments could further refine our understanding of effective engagement practices. For instance, discussions about text-heavy content differ in nature from those about charts or illustrations [18] as they require paraphrasing and summarization instead of describing or interpreting results, which may necessitate different curation strategies. Moreover, ensuring the inclusivity and accessibility of these strategies remains a critical concern. Future work should investigate how these interventions accommodate students with varying needs and learning preferences, ensuring that online discussion platforms are equitable and supportive for all participants. Lastly, the role of language and semantic choices in student engagement warrants further investigation. Our findings highlight the significant influence of wording on participation willingness, suggesting that a nuanced understanding of communication strategies could lead to more inclusive and engaging online learning communities.

## 6.3 Addressing Low-Quality Content in Mandatory Online Discussion Forum

Addressing the proliferation of low-quality content in online discussion forums, particularly where participation is mandated, represents a significant challenge in pursuing meaningful educational engagement. Mandatory participation often leads to contributions that are more about fulfilling requirements than fostering genuine discussion. To counteract this, the literature suggests social mechanisms like upvoting as a method to bring attention to high-quality contributions. However, our empirical observations and participant feedback reveal a notable reluctance towards using upvoting. This hesitation stems from concerns about potentially cluttering the discussion space and inadvertently placing undue emphasis on specific contributions, which could skew the perceived value of discussions and deter comprehensive engagement. Given these insights, our research has explored alternative strategies that circumvent the need for direct user actions, such as upvoting. A notable direction has been the adoption of instructor endorsements and the implementation of automated spotlighting mechanisms. These approaches are designed to elevate the visibility of contributions based on intrinsic content value or relevance to the community, including highlighting questions or content from individuals that users follow. This method aims to mitigate the issues associated with social upvoting by providing a more curated and less intrusive way of promoting quality discussions. Instructor endorsements, in particular, offer a nuanced form of highlighting content that leverages the authority and expertise of educators to signal the importance of specific discussions or contributions. This helps guide students toward valuable content and models the forum's substantive engagement criteria. On the other hand, automated spotlighting based on specific content criteria—such as the relevance of questions or the following relationships between users—presents a tailored approach to content promotion. By focusing on contributions' inherent value and personal relevance, this strategy aims to foster a more engaging and less cluttered discussion environment. These alternatives to upvoting are predicated on the understanding that quality engagement in online forums is multifaceted, requiring more than just a simple mechanism for elevating content. Instead, it demands a thoughtful consideration of how contributions are recognized and promoted, ensuring that the process enriches the discussion space without overwhelming it. By adopting these strategies, our work contributes to the broader discourse on enhancing online educational forums, offering insights into navigating the challenges of mandatory participation and low-quality content while promoting an inclusive and engaging learning environment.

## 7 FUTURE DIRECTION

In recent times, language models have undergone significant advancements, which have led to their diverse applications across various domains. This section explores the potential benefits these models can offer to enhance the user experience while using social annotation (SA) tools. In particular, we will examine the utility of AI, and more specifically, language models, during three key user activities: reading, writing annotations, and instructor-led spotlighting. By analyzing these activities, we can better understand how AI-powered language models can enhance the overall user experience and make social annotation tools more effective and efficient.

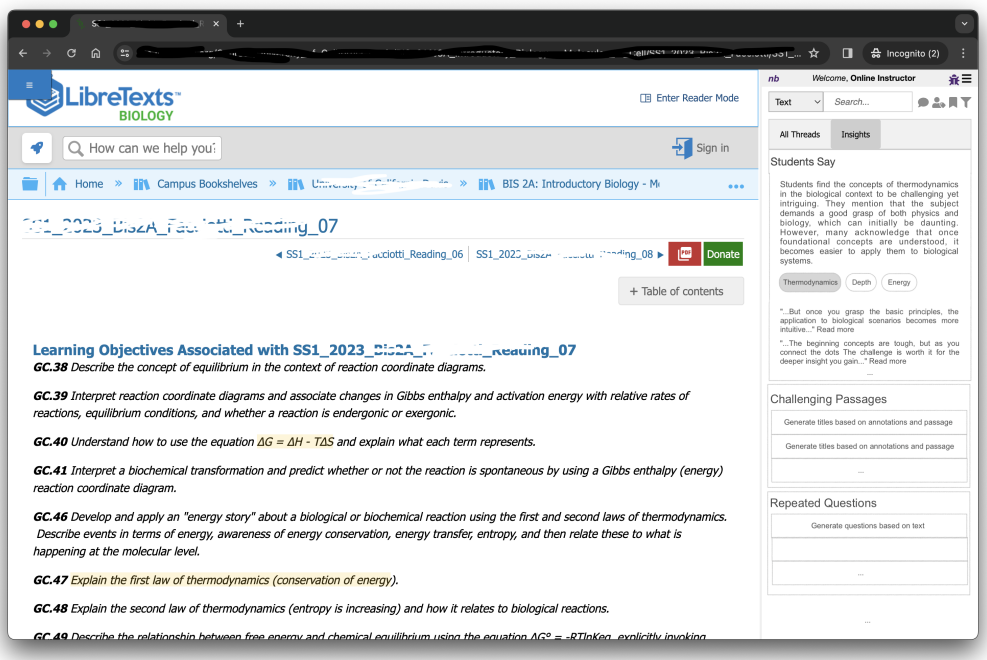

Fig. 4. Proposed Interface for Reading in Social Annotation Tools Using Language Models.

## 7.1 During Reading

Alongside the existing sidebar containing comments in the NB platform, students could switch to an "Insights" tab. This tab would feature a panel providing a summary of all student annotations generated by a language model. Students could navigate directly to specific annotations of interest, identified by frequently recurring keywords or those deemed critical within the reading material. A second panel might highlight challenging passages and synthesize corresponding annotations, utilizing both the original text and student contributions for context. To address redundancy, a common issue in large-scale discussions, a third panel could display the most frequently asked questions, with the language model generating queries based on common themes and grouping related student inquiries. This setup encourages student interaction, with system prompts designed to boost confidence and facilitate more meaningful exchanges. Figure 4 illustrates a proposed user interface for this enhanced reading experience.

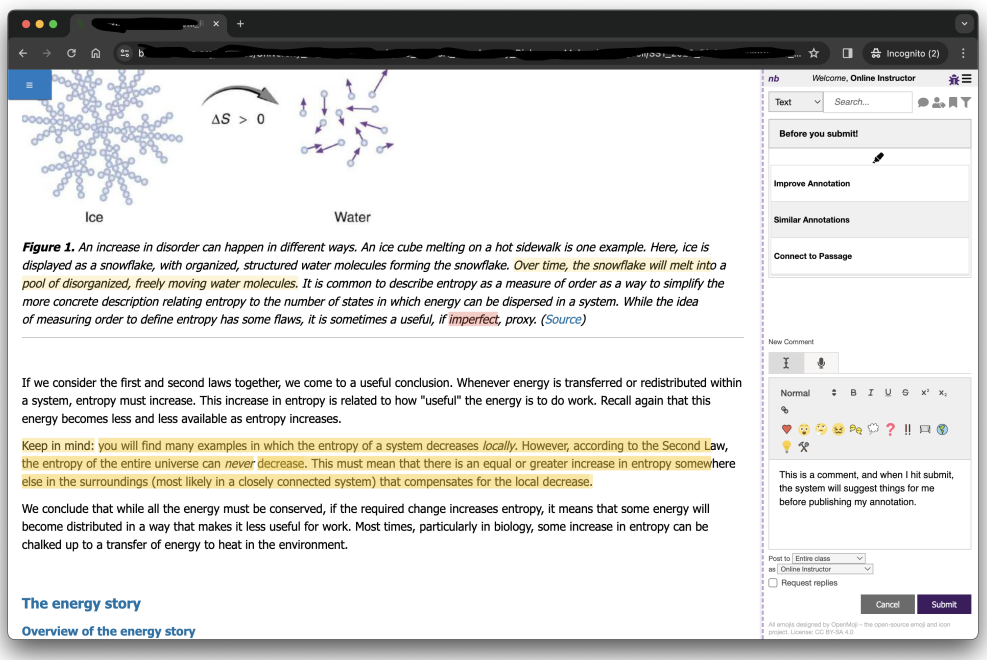

Fig. 5. Proposed Interface for composing annotations in Social Annotation Tools Using Language Models.

## 7.2 During Writing

Language models have the potential to assist students as they draft annotations. An interface can be designed to provide suggestions to students before they submit their comments, as illustrated in Figure 5. The suggestions can help students improve the quality of their content, meet grading criteria, refine their language use, and elaborate on their points. This feature can be particularly useful for international students who may require additional language support. Furthermore, the tool can alert students to existing annotations that are similar, encouraging them to either differentiate their comments or participate in existing discussions. This way, the tool reduces redundancy and promotes thoughtful participation.

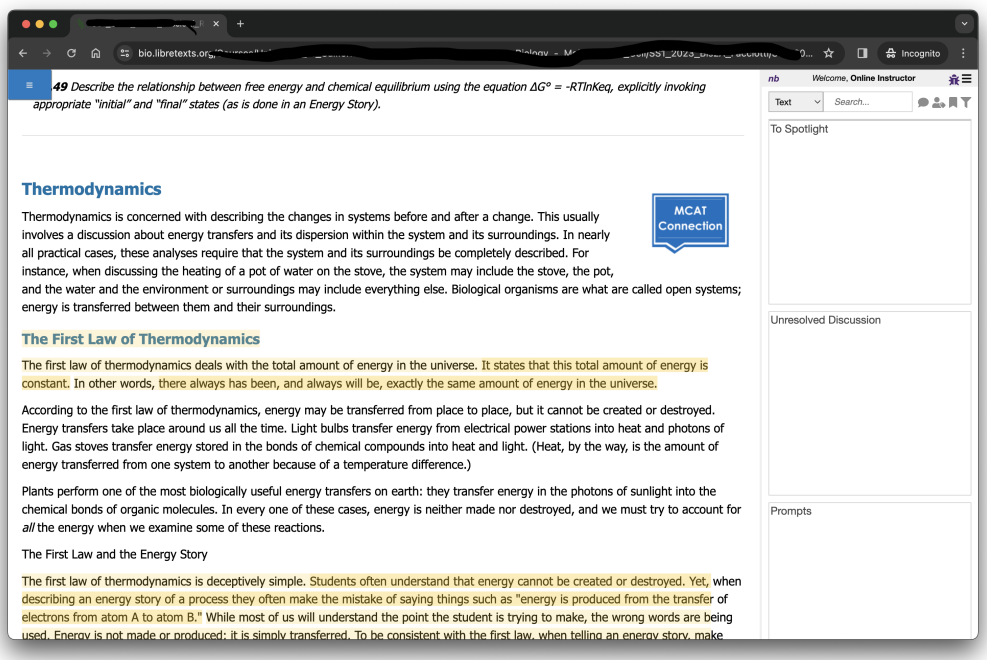

Fig. 6. Proposed Interface for instructors to spotlight annotations in Social Annotation Tools Using Language Models.

## 7.3 Instructor Spotlighting

According to our interviews, instructors value the personal connection with students gained through interacting with their annotations. However, managing thousands of comments can be overwhelming. Language models, informed by instructor-defined prompts and taxonomy, could suggest annotations for spotlighting, although final selections would remain at the instructor's discretion. An additional feature could highlight unresolved discussions, drawing instructor attention to potentially significant yet overlooked threads. This functionality not only aids in managing large volumes of data but also enhances the relevance and impact of instructor interventions. A sidebar could serve as a continuous interface for updating model prompts and validating outputs, ensuring alignment with instructional goals (see Figure 6).